\documentclass[onecolumn,showpacs,preprintnumbers,superscriptaddress,amsmath,amssymb]{revtex4}
\usepackage[dvips]{graphicx}
\usepackage{latexsym,amssymb,amsmath,epsfig,bm,times,psfrag}
\usepackage{color}
\usepackage[bookmarksnumbered,bookmarksopen,colorlinks,citecolor=red,linkcolor=blue]{hyperref}
\usepackage{epstopdf}
\usepackage[capitalize]{cleveref}
\newcommand{\Fig}[1]{Fig.~\ref{#1}}
\newcommand{\eq}[1]{Eq.~(\ref{#1})}
\newcommand{\sect}[1]{Sec.~\ref{#1}}

\renewcommand{\part}{{\rm part}}

\renewcommand{\vec}{\boldsymbol}
\newcommand{\be}{\begin{equation}}
\newcommand{\ee}{\end{equation}}
\newcommand{\bear}{\begin{eqnarray}}
\newcommand{\eear}{\end{eqnarray}}
\newcommand{\ba}{\begin{array}}
\newcommand{\ea}{\end{array}}

\begin{document}

\title{Quark condensate and magnetic moment in a strong magnetic field}

\author{De-Xian Wei}
\affiliation{School of Science, Guangxi University of Science and Technology, Liuzhou, 545006, China}

\author{Li-Juan Zhou}
\email{zhoulj@gxust.edu.cn}
\affiliation{School of Science, Guangxi University of Science and Technology, Liuzhou, 545006, China}
\date{\today}

\begin{abstract}
This paper studies the quark condensate, magnetic moment, magnetic polarization, and magnetic susceptibility
in a strong external magnetic field by employing the Dyson-Schwinger equations (DSE).
The results show that these physical quantities as functions of the magnetic field.
We note that the quark's spin polarizations are approximately proportional to the magnetic field magnitude.
For comparison, we investigate the magnetic moments and susceptibility of the nucleon
in the constituent quark model framework and demonstrate that both these quantities increase as the magnetic field rises.
\end{abstract}


\maketitle


\section{Introduction}
\label{sec:sec1}

Studying the QCD phase structure is always a hot research topic in high-energy physics.
Many physical parameters affect the QCD phase diagram, where besides the temperature and chemical potential,
and also the external magnetic field affect the diagram. Since quarks are electrically charged
and can be coupled to magnetic field, the latter can affect the phase structure of QCD matter.
Recently, with the development of astronomical observation and heavy-ion collision experiments,
very strong magnetic field have been generated in both astrophysical objects and noncentral heavy-ion collision experiments.
Currently, the strengths of the magnetic field on the astrophysical objects called magnetars
are believed to be of the order of $10^{-4}$~GeV$^{2}$~\cite{Duncan:1992fov}.
Moreover, the generated magnetic field is about $eB\sim 0.1$~GeV$^{2}$ at RHIC while $eB\sim 1$~GeV$^{2}$ at LHC~\cite{Deng:2012ego}.
Due to the asymptotic freedom feature of QCD, the QCD matter will undergo a phase transition
from the hadronic phase to the quark-gluon plasma phase. Such a strong magnetic field significantly influences
the behavior of charged particles produced in QCD matter, and new QCD phase structures may appear.
The physics of QCD and related phenomena has been reviewed
in Refs.~\cite{Miransky:2015qft,Andersen:2016pdo} which contain an extensive set of references.

Theoretical physicists have put forward many theoretical models to study the influence of magnetic field
on strongly interacting systems, including the lattice QCD~\cite{Elia:2010qpt,Bali:2012mso,Bonati:2014msa,Ding:2022cca}
and the Nambu-Jona-Lasinio (NJL) model~\cite{Gatto:2013qmi,Dumm:2017smf,Fraga:2014afi,Kawaguchi:2022rot}, etc.
The experimental results revealed that with the presence of a magnetic field background,
the strongly interacting matter presents many exotic phenomena in QCD phase structure,
e.g., the chiral separation induced by the magnetic field which noted as Chiral Magnetic Effect (CME)~\cite{Kharzeev:2008teo},
the magnetic dimensional reduction induces the enhancement of the chiral symmetry breaking in the vacuum
is notes as Magnetic Catalysis (MC)~\cite{Klevansky:1989csr,Gusynin:1996dra,Andersen:2012cpt,Bali:2012tqp,Bali:2012qqc,Bali:2013mfg},
the critical temperature of chiral phase transition decreases with the magnetic field
is called for the Inverse Magnetic Catalysis (IMC)~\cite{Bali:2012tqp,Bali:2013mfg,Fayazbakhsh:2014amm,Chaudhuri:2020eoq,Chaudhuri:2022apo,Bandyopadhyay:2021imc}.

In our previous work, we studied the properties and structure of the QCD vacuum using
the Dyson-Schwinger equations (DSE) of the quark propagator at zero and finite temperature.
Then we calculated the various condensation values of the quark and gluons,
the quark-gluon mixed condensation values and quark effective mass~\cite{Ma:2012tpo,Zhou:2014tdo,Zhou:2010nms}.
Recently studies~\cite{Elia:2010qpt,Bali:2012mso,Bonati:2014msa,Gatto:2013qmi,Dumm:2017smf, Fraga:2014afi,Zhang:2015eos,Zhang:2016pom,Fu:2017fia,Shi:2015dcs,Liu:2018nac}
revealed that the properties and condensation values of QCD vacuum might change in strong magnetic field.
Therefore, this paper extends the previous work in strong magnetic field.
Specifically, we follow the work of Mueller and his collaborators~\cite{Mueller:2014dqm},
regarding the quark propagator in strong magnetic field at zero temperature,
and use the DSE to study the properties of QCD vacuum, the subtracted quark condensates,
the $eB$ dependence of the magnetic moment, spin polarization, and magnetic susceptibility.
Based on the constituent quark model, we also study the $eB$ dependence of magnetic moments and magnetic susceptibility of the nucleon
and try to understand the influence of strong magnetic field on the phase transition of QCD.
This study provides theoretical guidance for searching quark-gluon plasma
in relativistic heavy-ion collision experiments.

The remainder of this paper is organized as follows.
\sect{sec:sec2} briefly describes the basic theory of DSE in a strong external magnetic field.
\sect{sec:sec3} presents the numerical results about the quark condensates,
magnetic moments, magnetic polarization, spin polarizations, and magnetic susceptibility.
Finally, \sect{sec:sum} summarizes the main results.

\section{Theory model}
\label{sec:sec2}
This section reviews the DSE formula, which is widely used in
the non-perturbative region of QCD and in some other fields like the Quantum Electrodynamics in (2+1) dimensions (QED3)~\cite{Li:2014tcp}, etc.

The DSE in position space and with local interaction is given by
\begin{eqnarray}\label{dse:def101}
S^{-1}(x,y) &=& S_{0}^{-1}(x,y)+\Sigma(x,y),
\end{eqnarray}
where $S^{-1}$ is the inverse of dressed quark propagator,
$S_{0}^{-1}$ is the inverse of free quark propagator, and the quark self-energy reads
\begin{eqnarray}\label{dse:def102}
\Sigma(x,y) &=& ig^{2}C_{F}\gamma^{\mu}S(x,y)\Gamma^{\nu}(y)D_{\mu\nu}(x,y),
\end{eqnarray}
with $g$ is the coupling constant of a strong interaction,
and $\Gamma^{\nu}$ is the dressed quark-gluon vertex.
The Casimir $C_{F} = (N_{c}^{2}-1)/N_{c}$ stems from
the colour trace [$C_{F}\delta_{ij}=(T^{a}T^{a})_{ij}$,
$T$ is the $SU$(3) generators in the fundamental representation].
Furthermore, $D_{\mu\nu}$ denotes the gluon propagator in Landau gauge.
We expand \eq{dse:def101} in terms of Ritus transformation functions, i.e.,
multiplying this equation with $\bar{E}_{p}(x)$ from the left and $E_{p^{'}}(y)$ from the right.
The integration over $x$ and $y$ yields~\cite{Ritus:1978moe,Ritus:1987iqe}
\begin{eqnarray}\label{dse:def105}
\int d^{4}xd^{4}y\bar{E}_{p}(x)S^{-1}(x,y)E_{p^{'}}(y) &=&  \int d^{4}xd^{4}y\bar{E}_{p}(x)S_{0}^{-1}(x,y)E_{p^{'}}(y) +\int d^{4}xd^{4}y\bar{E}_{p}(x)\Sigma(x,y)E_{p^{'}}(y).
\end{eqnarray}
Using
\begin{eqnarray}\label{dse:def106}
\int d^{4}x\bar{E}_{p}(x)E_{p^{'}}(x) &=& (2\pi)^{4}\delta^{4}(p-p^{'})\Pi(L), \nonumber\\
\sum_{L=0}^{\infty}\int\frac{d^{2}p_{\parallel}}{(2\pi)^{4}}\int_{-\infty}^{\infty}dp_{2}E_{p}(x)\bar{E}_{p}(y) &=& (2\pi)^{4}\delta^{4}(x-y),
\end{eqnarray}
and
\begin{eqnarray}\label{dse:def107}
 \Pi(L)=
   \begin{cases}
        \Delta(sgn(eB)), & L=0,\\
        1,  &  L>0   ,
    \end{cases}
\end{eqnarray}
one obtains
\begin{eqnarray}\label{dse:def108}
(2\pi)^{4}\delta^{4}(p-p^{'})\Pi(L)\left[A_{0}(p)+A_{\parallel}(p)i\gamma\cdot p_{\parallel}+A_{\perp}(p)i\gamma\cdot p_{\perp}\right] &=&  (2\pi)^{4}\delta^{4}(p-p^{'})\Pi(L)\left[\gamma p+m\right]+\Sigma(p,p^{'}),
\end{eqnarray}
where $A_{0}$ is the scalar quark dressing function of the quark propagator in (pseudo-)momentum space,
$A_{\parallel}$ and $A_{\perp}$ are vector quark dressing functions of the quark propagator
in (pseudo-)momentum space, whereas $\Sigma(p,p^{'})$ denotes the self-energy.
The momentum vectors parallel and perpendicular to the magnetic field direction are
denoted by $p_{\parallel}=(p_{0},~0,~0,~p_{3})^{T}$ and $p_{\perp}=(0,~0,~p_{2}=\sqrt{2|eB|L},~0)^{T}$
($eB$ is the magnitude of magnetic field, $L$ is the total angular momentum quantum for each Landau level).
The self-energy term is implicitly proportional to $\delta^{4}(p-p^{'})\Pi(L)$.
The quark self-energy in the Ritus eigenbasis is then given by
\begin{eqnarray}\label{dse:def109}
\Sigma(p,p^{'}) &=&  g^{2}C_{F}\int d^{4}xd^{4}y\bar{E}_{p}(x)\gamma^{\mu}S(x,y)\Gamma^{\nu}(y)D_{\mu\nu}(x,y)E_{p^{'}}(y),
\end{eqnarray}
where $S(x,y)$ is the fermion propagator. The representation of fermion propagator in Ritus's representation is given by
\begin{eqnarray}\label{dse:def110}
S(x,y) &=&  \sum_{L=0}^{\infty}\int\frac{d^{2}p_{\parallel}}{(2\pi)^{4}}\int_{-\infty}^{\infty}dp_{2} E_{q}(x)\frac{1}{A_{0}(q)+i\gamma\cdot q_{\parallel}A_{\parallel}(q)+i\gamma\cdot q_{\perp}A_{\perp}(q)}\bar{E}_{q}(y),
\end{eqnarray}
with scalar and vector quark dressing functions $A_{0}$, $A_{\parallel}$, $A_{\perp}$.
For the bare quark propagator $S_{0}$ we have $A_{\parallel}=A_{\perp}$=1 and $A_{0}=Z_{m}m$, with bare quark mass $m$.
The idea behind the Ritus method is to use the Ritus transformation functions $E_{q}$ and $\bar{E}_{q}$,
as a substitute for the usual Fourier exponential factor $e^{ip\cdot x}$.
It is worthy note that by Ritus's work, the fermion two-point Green's function solely depends on four independent Lorentz structures,
i.e., $\gamma\Pi$, $\sigma F$, $(F\Pi)^{2}$ and $\gamma^{5}FF^{*}$~\cite{Ritus:1978moe,Ritus:1987iqe}.
In principle, due to the appearance of further Lorenz structures ($\propto F_{\mu\nu}$),
the fermion propagator could possess a richer tensor structure~\cite{Ferrer:2009diz,Ferrer:2010dga,Ferrer:2014nla}.
However, any additional spin-dependent tensor structures violate a remaining $Z(2)$ symmetry of the system
by rendering the position of a putative pole structure in the quark propagator
dependent on the direction of the external field~\cite{Leung:2007gia,Leung:2006gia}.
Consequently, given the additional structure, it is possible to obtain non-trivial solutions
for $A_{\parallel}$, $A_{\perp}$, and $A_{0}$ only together with zero dressing functions in these additional structures.
Therefore, we only consider concise structures of two-point functions in this work.

To evaluate \eq{dse:def109}, we must know $D_{\mu\nu}(x,y)$. The isotropic Fourier representation of the Landau gauge gluon is given by
\begin{eqnarray}\label{dse:def111}
D_{\mu\nu}(x,y)=\int \frac{d^{4}k}{(2\pi)^{4}}e^{ik(x-y)}D(k^{2})P_{\mu\nu},
\end{eqnarray}
where $D(k^{2})$ is gluon propagator function with its explicit form well be shown as following,
and $P_{\mu\nu}=\delta_{\mu\nu}-k_\mu k_\nu/k^2$ is the transverse projector.
By substituting \eq{dse:def110} and \eq{dse:def111} into \eq{dse:def109}, one obtains
\begin{eqnarray}\label{dse:def112}
\Sigma(p,p^{'}) &=& g^{2}C_{F} \sum_{L=0}^{\infty}\int\frac{d^{2}p_{\parallel}}{(2\pi)^{4}}\int_{-\infty}^{\infty}dp_{2}\int\frac{d^{4}k}{(2\pi)^{4}} \int d^{4}xd^{4}y \nonumber\\
&\times& \bar{E}_{p}(x)\gamma^{\mu}E_{q}(x)\frac{1}{\left[A_{0}(q)+A_{\parallel}(q)i\gamma\cdot q_{\parallel}+A_{\perp}(q)i\gamma\cdot q_{\perp}\right]}\bar{E}_{q}(y)\Gamma^{\nu}E_{p^{'}}(y)e^{ik(x-y)}D(k^{2})P_{\mu\nu}.
\end{eqnarray}

The dressed quark-gluon vertex $\Gamma^{\nu}$ is a much more difficult object,
which is unknown in detail, even in the case of vanishing background fields.
To make the equations tractable, we resort to a simple ansatz of the form $\Gamma^{\nu} \rightarrow \gamma^{\nu}\Gamma(k^{2})$
based on Ref.~\cite{Mueller:2014dqm}, where $\Gamma(k^{2})$ is considered independent of the magnetic field,
and its explicit form is given in \eq{dse:def213}.

After calculating the integral over $x$ and $y$, and involving a product of Ritus and Fourier eigenfunction, the quark self-energy equation is given~\cite{Mueller:2014dqm}
\begin{eqnarray}\label{dse:def202}
\Sigma(p,p^{'}) &=&  (2\pi)^{4}\delta^{3}(p-p^{'})ig^{2}C_{F}\sum_{L_{q}=0}^{\infty}\int\frac{d^{2}q_{\parallel}}{(2\pi)^{4}}\int_{-\infty}^{\infty}dq_{2}\int_{-\infty}^{\infty}dk_{1} e^{-\frac{k_{\perp}^{2}}{|2eB|}} \sum_{\sigma_{1}\sigma_{2}\sigma_{3}\sigma_{4}}\delta_{n(\sigma_{1},L)}\delta_{n(\sigma_{2},L_{q})}\delta_{n(\sigma_{3},L_{q})}\delta_{n(\sigma_{4},L^{'})}  \nonumber\\
&\times&  \triangle(\sigma_{1})\gamma^{\mu}\triangle(\sigma_{2})\frac{1}{\left[A_{0}(q)+A_{\parallel}(q)i\gamma\cdot q_{\parallel}+A_{\perp}(q)i\gamma\cdot q_{\perp}\right]}\triangle(\sigma_{3})\gamma^{\nu}\triangle(\sigma_{4})D(k^{2})\Gamma(k^{2})P^{\mu\nu}(k).
\end{eqnarray}

After performing the traces in the quark DSE, we obtain the dressing functions, which are given by~\cite{Mueller:2014dqm}
\begin{eqnarray}\label{dse:def203}
A_{0}(p)|_{L_{p}=L} &=& Z_{2}m +C_{1}\int_{q}\left\{\left(\frac{A_{0}(q)}{A_{0}^{2}(q)+A_{\parallel}^{2}(q)q_{\parallel}^{2}+ A_{\perp}^{2}(q)q_{\perp}^{2}}\right)\Bigg|_{L_{q}=L} \cdot e^{-\frac{k_{\perp}^{2}}{|2eB|}}G_{1}(k^{2})D(k^{2})\Gamma(k^{2})\right\} ~\nonumber\\
 &+& \frac{C_{2}}{p_{\parallel}^{2}}\frac{2}{\tau(L)}\sum_{L_{q}=L\pm1}\int_{q}\left\{\left(\frac{A_{0}(q)}{A_{0}^{2}(q)+A_{\parallel}^{2}(q)q_{\parallel}^{2}+ A_{\perp}^{2}(q)q_{\perp}^{2}}\right)\Bigg|_{L_{q}}
 \cdot e^{-\frac{k_{\perp}^{2}}{|2eB|}}G_{2}(k^{2})D(k^{2})\Gamma(k^{2})\right\},
\end{eqnarray}

\begin{eqnarray}\label{dse:def204}
A_{\parallel}(p)|_{L_{p}=L} &=& Z_{2} -\frac{C_{1}}{p_{\parallel}^{2}}\int_{q}
 \left\{\left(\frac{A_{\parallel}(q)}{A_{0}^{2}(q)+A_{\parallel}^{2}(q)q_{\parallel}^{2}+ A_{\perp}^{2}(q)q_{\perp}^{2}}\right)\Bigg|_{L_{q}=L}\cdot e^{-\frac{k_{\perp}^{2}}{|2eB|}}G_{3}(p,q,k^{2})D(k^{2})\Gamma(k^{2})\right\} ~\nonumber\\
 &+& \frac{C_{2}}{p_{\parallel}^{2}}\frac{2}{\tau(L)}\sum_{L_{q}=L\pm1}\int_{q}\left\{\left(\frac{A_{\parallel}(q)}{A_{0}^{2}(q)+A_{\parallel}^{2}(q)q_{\parallel}^{2}+ A_{\perp}^{2}(q)q_{\perp}^{2}}\right)\Bigg|_{L_{q}}\cdot e^{-\frac{k_{\perp}^{2}}{|2eB|}}G_{4}(p,q,k^{2})D(k^{2})\Gamma(k^{2})\right\},
\end{eqnarray}

\begin{eqnarray}\label{dse:def205}
A_{\perp}(p)|_{L_{p}=L} &=& Z_{2} +\frac{C_{1}}{p_{\parallel}^{2}}\int_{q}
 \left\{\left(\frac{A_{\perp}(q)}{A_{0}^{2}(q)+A_{\parallel}^{2}(q)q_{\parallel}^{2}+ A_{\perp}^{2}(q)q_{\perp}^{2}}\right)\Bigg|_{L_{q}=L}\cdot e^{-\frac{k_{\perp}^{2}}{|2eB|}}G_{5}(p,q,k^{2})D(k^{2})\Gamma(k^{2})\right\} ~\nonumber\\
 &-& \frac{C_{2}}{p_{\parallel}^{2}}\frac{2}{\tau(L)}\sum_{L_{q}=L\pm1}\int_{q}\left\{\left(\frac{A_{\perp}(q)}{A_{0}^{2}(q)+A_{\parallel}^{2}(q)q_{\parallel}^{2}+ A_{\perp}^{2}(q)q_{\perp}^{2}}\right)\Bigg|_{L_{q}}\cdot e^{-\frac{k_{\perp}^{2}}{|2eB|}}G_{6}(p,q,k^{2})D(k^{2})\Gamma(k^{2})\right\},
\end{eqnarray}
where
\begin{eqnarray}\label{dse:def206}
   \begin{cases}
        \tau(L)=2, & L=0\\
        \tau(L)=4, & L>0   .
    \end{cases}
\end{eqnarray}
We make the following identifications in \eq{dse:def203}-\eq{dse:def205}
\begin{eqnarray}\label{dse:def211}
\int_{q} &=& \int\frac{d^{2}q_{\parallel}}{(2\pi)^{4}}\int_{-\infty}^{\infty}dq_{2}dk_{1}, ~\nonumber\\
C_{1} &=& Z_{1f}g^{2}C_{F}, ~~~~ C_{2} = g^{2}C_{F}, ~\nonumber\\
G_{1}(k^{2}) &=& 2-\frac{k_{\parallel}^{2}}{k^{2}}, ~~~~ G_{2}(k^{2})= 2-\frac{k_{\perp}^{2}}{k^{2}}, ~\nonumber\\
G_{3}(p,q,k^{2}) &=& p_{\parallel}q_{\parallel}cos(\varphi)\frac{k_{\parallel}^{2}}{k^{2}} -\frac{2[p_{\parallel}q_{\parallel}cos(\varphi)-p_{\parallel}^{2}][q_{\parallel}^{2}-p_{\parallel}q_{\parallel}cos(\varphi)]}{k^{2}}, ~\nonumber\\
G_{4}(p,q,k^{2}) &=& \left(2-\frac{k_{\perp}^{2}}{k^{2}}\right)p_{\parallel}q_{\parallel}cos(\varphi), ~\nonumber\\
G_{5}(p,q,k^{2}) &=& \left(2-\frac{k_{\parallel}^{2}}{k^{2}}\right)p_{\perp}q_{\perp}, ~\nonumber\\
G_{6}(p,q,k^{2}) &=& \left(\frac{k_{1}^{2}-k_{2}^{2}}{k^{2}}\right)p_{\perp}q_{\perp}, ~\nonumber\\
cos(\varphi) &=& \frac{\vec{p}_{\parallel}\vec{q}_{\parallel}}{|p_{\parallel}||q_{\parallel}|}.  ~\nonumber
\end{eqnarray}

Solving DSE in a strong magnetic field is very challenging. Therefore, for simplicity,
we choose the following forms for the quenched gluon propagator $D(k^{2})$ and the vertex dressing function~\cite{Fischer:2010cad}:
\begin{eqnarray}\label{dse:def212}
D(k^{2}) &=& \frac{1}{k^{2}}\frac{k^{2}\Lambda^{2}}{(k^{2}+\Lambda^{2})^{2}} \left\{\left(\frac{c}{k^{2}+a\Lambda^{2}}\right)^{b} +\frac{k^{2}}{\Lambda^{2}}\left[\frac{\beta\alpha(\mu)log[\frac{k^{2}}{\Lambda^{2}}+1]}{4\pi}\right]^{\gamma}\right\},
\end{eqnarray}

\begin{eqnarray}\label{dse:def213}
\Gamma(k^{2}) &=& \frac{d_{1}}{d_{2}+k^{2}}+\frac{k^{2}}{k^{2}+\Lambda^{2}}\left[\frac{\beta\alpha(\mu)log[\frac{k^{2}}{\Lambda^{2}}+1]}{4\pi}\right]^{2\delta}, 
\end{eqnarray}
where the parameters are $a$=0.60,~$b$=1.36,~$\Lambda$=1.4~GeV,~$c$=11.5~GeV$^{2}$,~$\beta$=22/3,~$\gamma$=-13/22, ~$d_{1}$=7.9~GeV$^{2}$,~$d_{2}$=0.5~GeV$^{2}$,~$\delta$=-18/88.
The quark mass renormalization factor $Z_{2}$ is determined in the renormalisation process.
The renormalization factor of the quark-gluon vertex is denoted by $Z_{1f}$, satisfying an approximate Slavnov-Taylor identity
in the infrared and the correct ultraviolet running from the resummed perturbation theory.

Substituting \eq{dse:def212}-\eq{dse:def213} into \eq{dse:def203}-\eq{dse:def205},
one can solve the equations numerically, and get the quark dressing function $A_{0}$, $A_{\parallel}$ and $A_{\perp}$.
Note that only $A_{0}$ and $A_{\parallel}$ contribute to the lowest Landau level.
More thorough discussions of the DSE can be found in Ref.~\cite{Mueller:2014dqm}.

To study the quark condensate and magnetization phenomena in the external magnetic field,
we use the finite quarks mass, i.e., $m_{u}=6$~MeV, $m_{d}=10$~MeV, $m_{s}=199$~MeV, in simulations.
Throughout this paper, our results
and analyses are considered on zero temperature ($T=0$).

\section{Numerical Results}
\label{sec:sec3}

\begin{figure*}[tp]
\begin{center}
\includegraphics[width=0.50\textwidth]{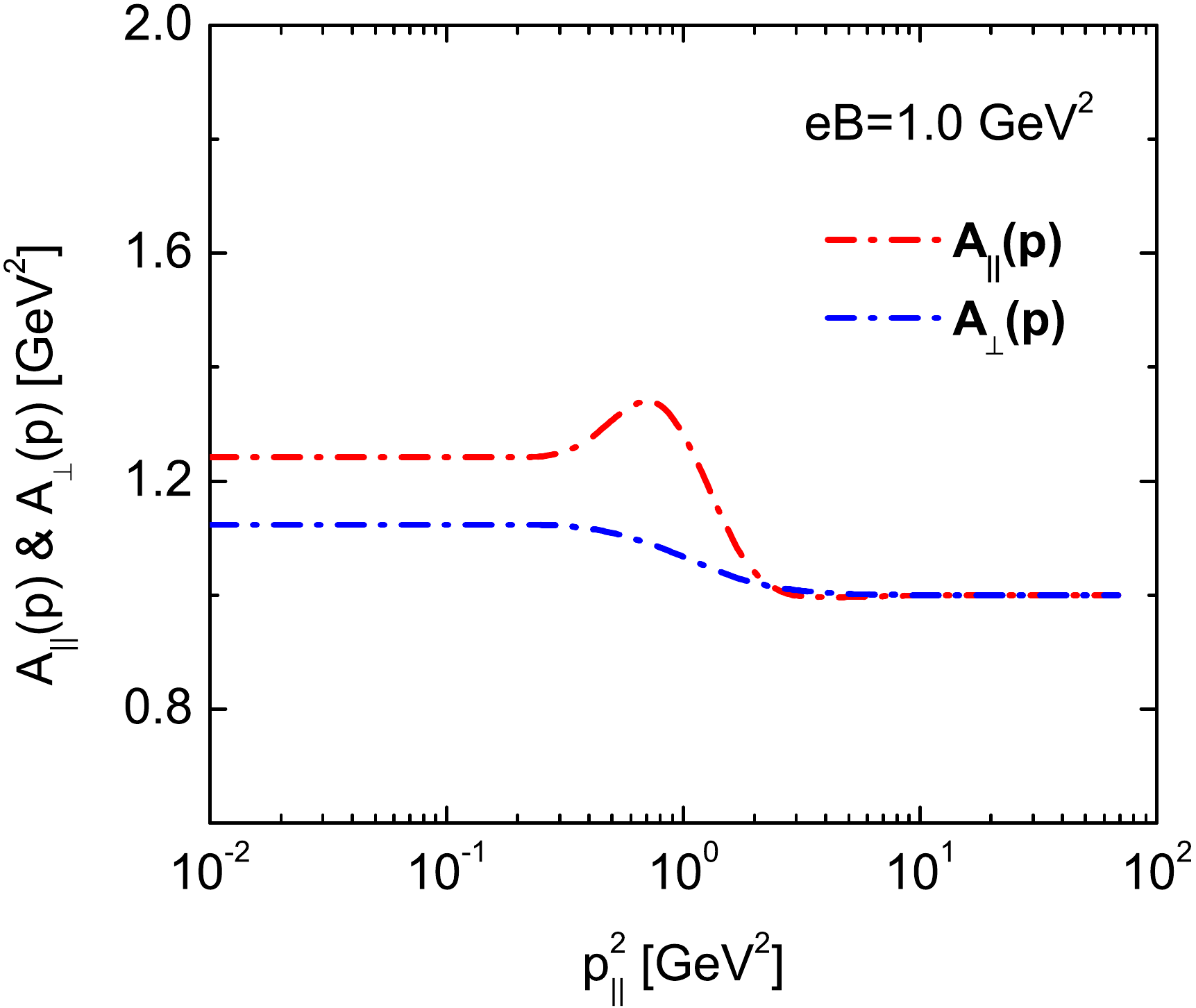}
\caption{(Color online)
The quenched dressing functions $A_{\parallel}$ and $A_{\perp}$ (L=1), as functions of the momentum $p_{\parallel}^{2}$ for $m_{u}$ = 6 MeV at $eB$=1.0~GeV$^{2}$.
}
\label{fig1}
\end{center}
\end{figure*}

\begin{figure*}[tp]
\begin{center}
\includegraphics[width=0.50\textwidth]{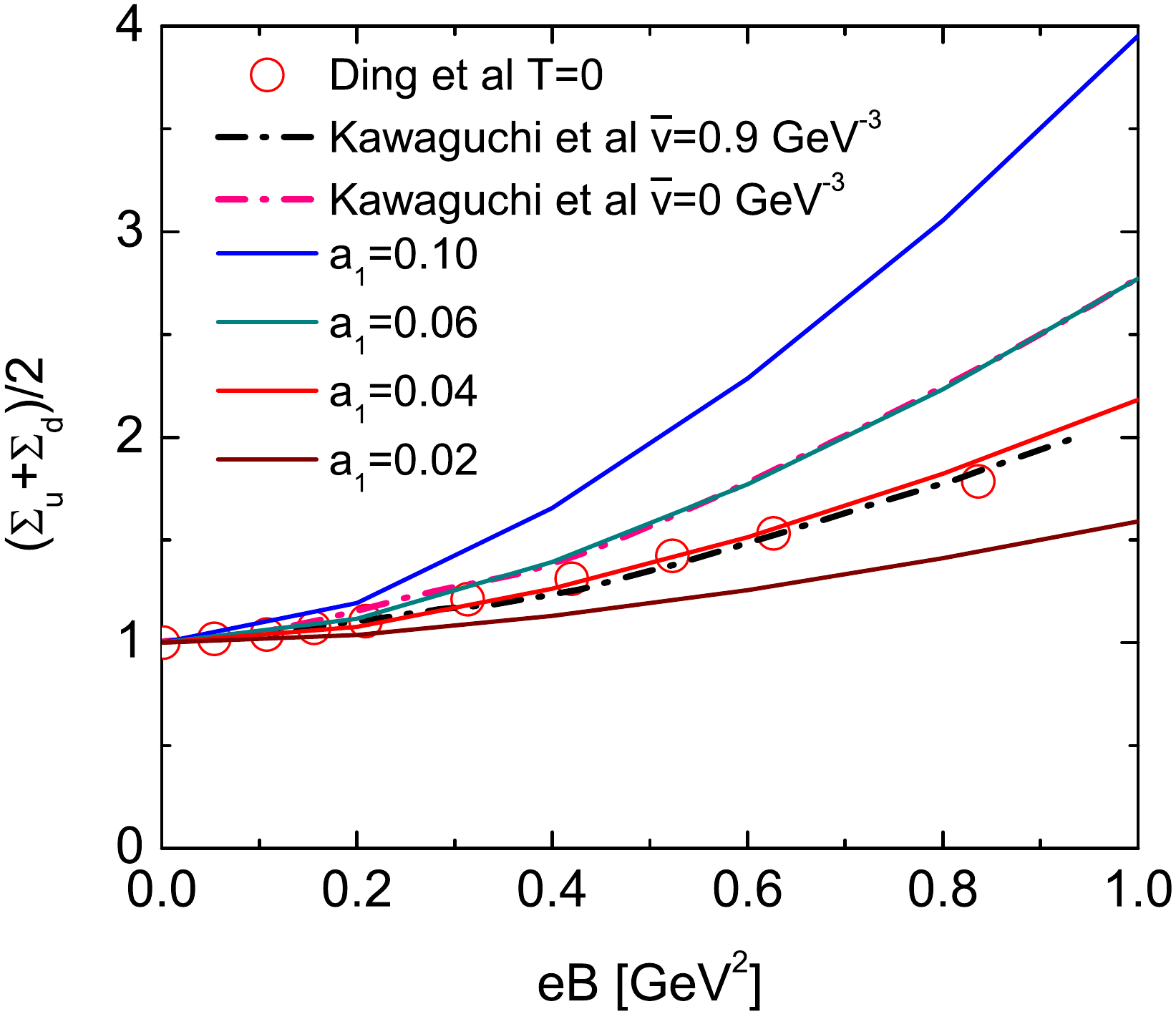}
\caption{(Color online)
The subtracted quark condensate, as a function of the magnetic field $eB$. Comparisons with our simulations (solid lines), the NJL [with AMM ($\kappa_{u}=\kappa_{d}=\bar{v}\sigma^{2}=0.9\sigma^{2}$~GeV$^{-1}$) and without AMM ($\kappa_{u}=\kappa_{d}=\bar{v}\sigma^{2}=0$~GeV$^{-1}$)] results (dash dot lines)~\cite{Kawaguchi:2022rot}, and the lattice QCD results (circle points)~\cite{Ding:2022cca} at $T=0$.
}
\label{fig2}
\end{center}
\end{figure*}

\begin{figure*}[tp]
\begin{center}
\includegraphics[width=0.320\textwidth]{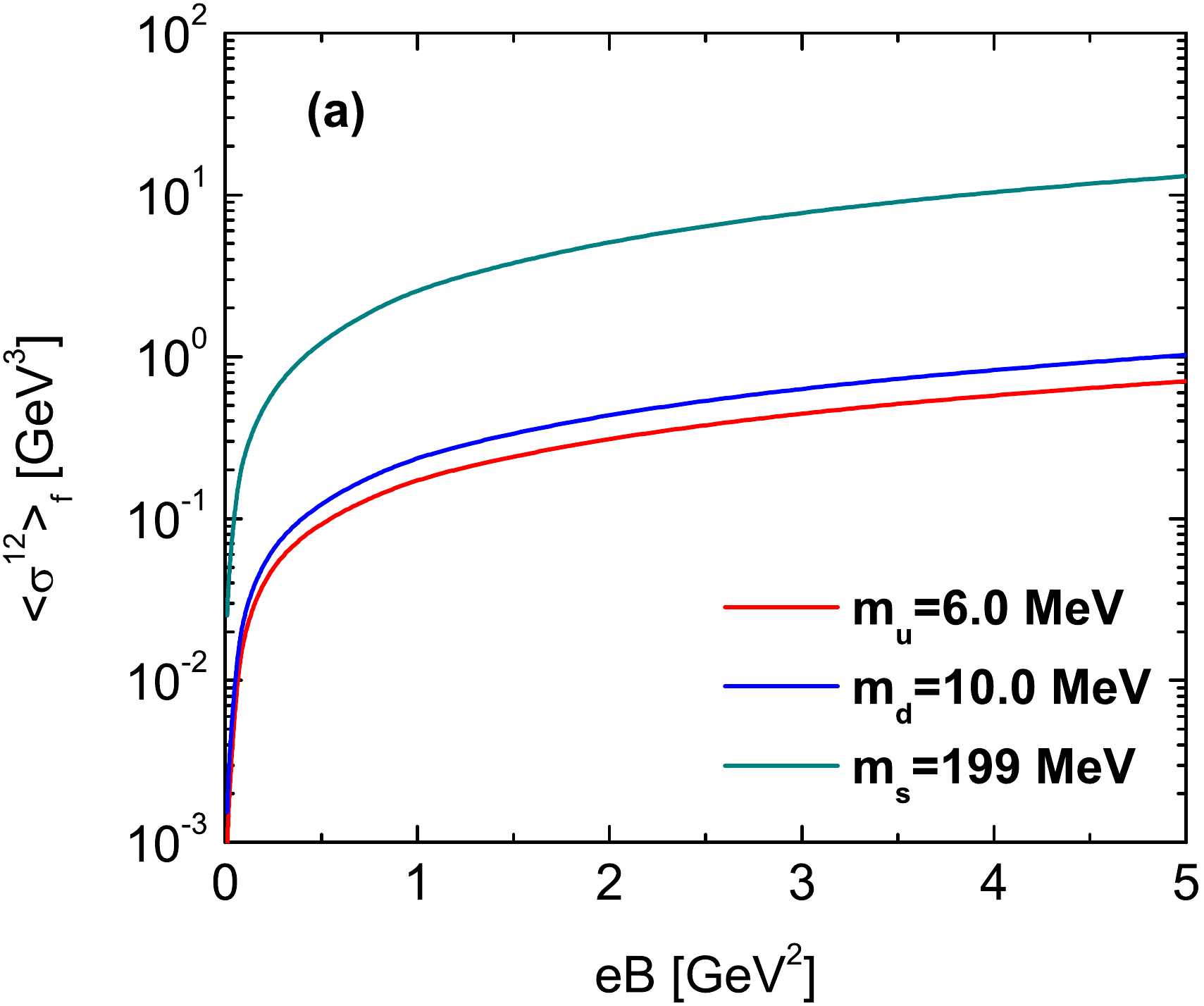}
\hspace{0.25cm}
\includegraphics[width=0.320\textwidth]{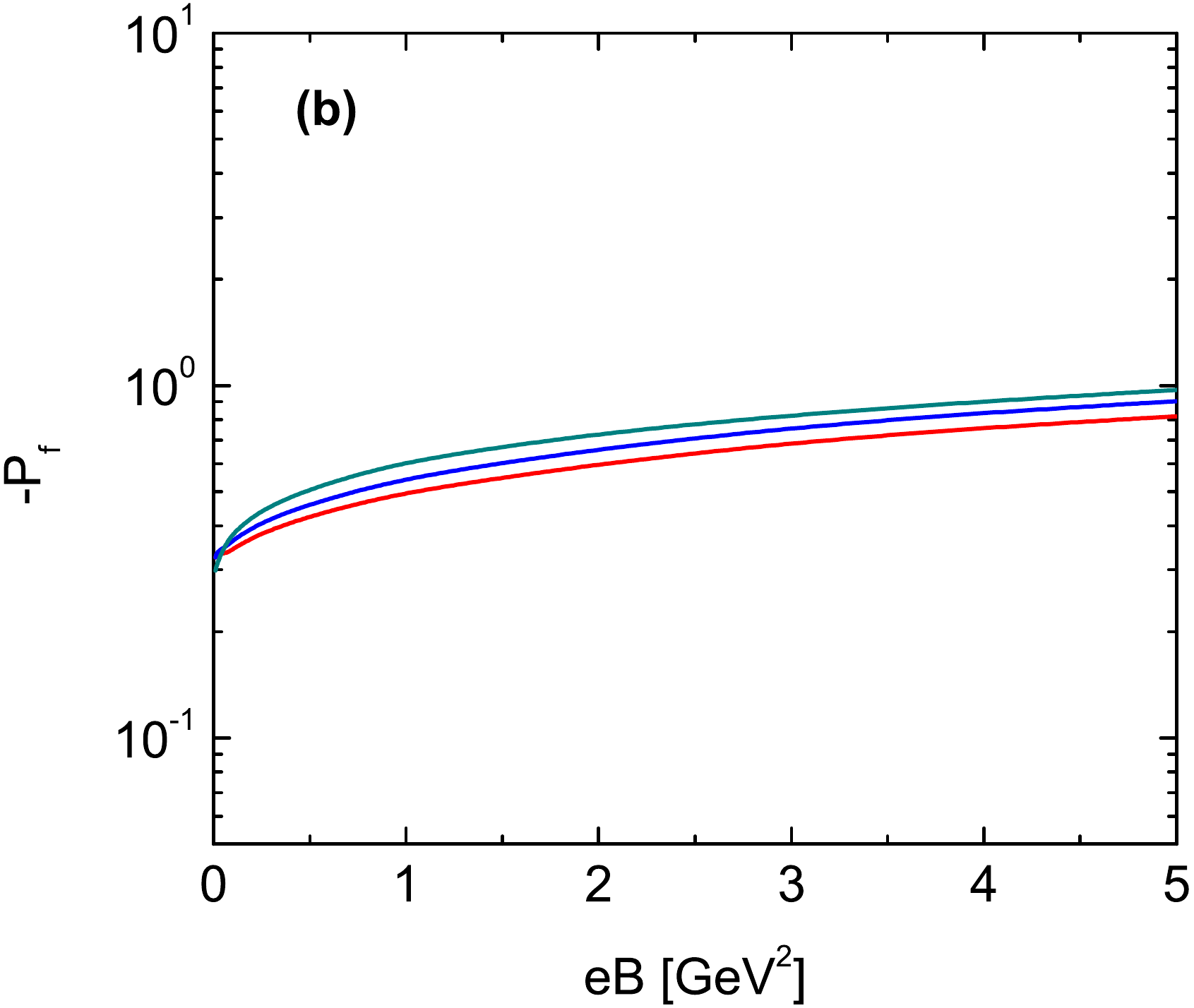}
\hspace{0.25cm}
\includegraphics[width=0.300\textwidth]{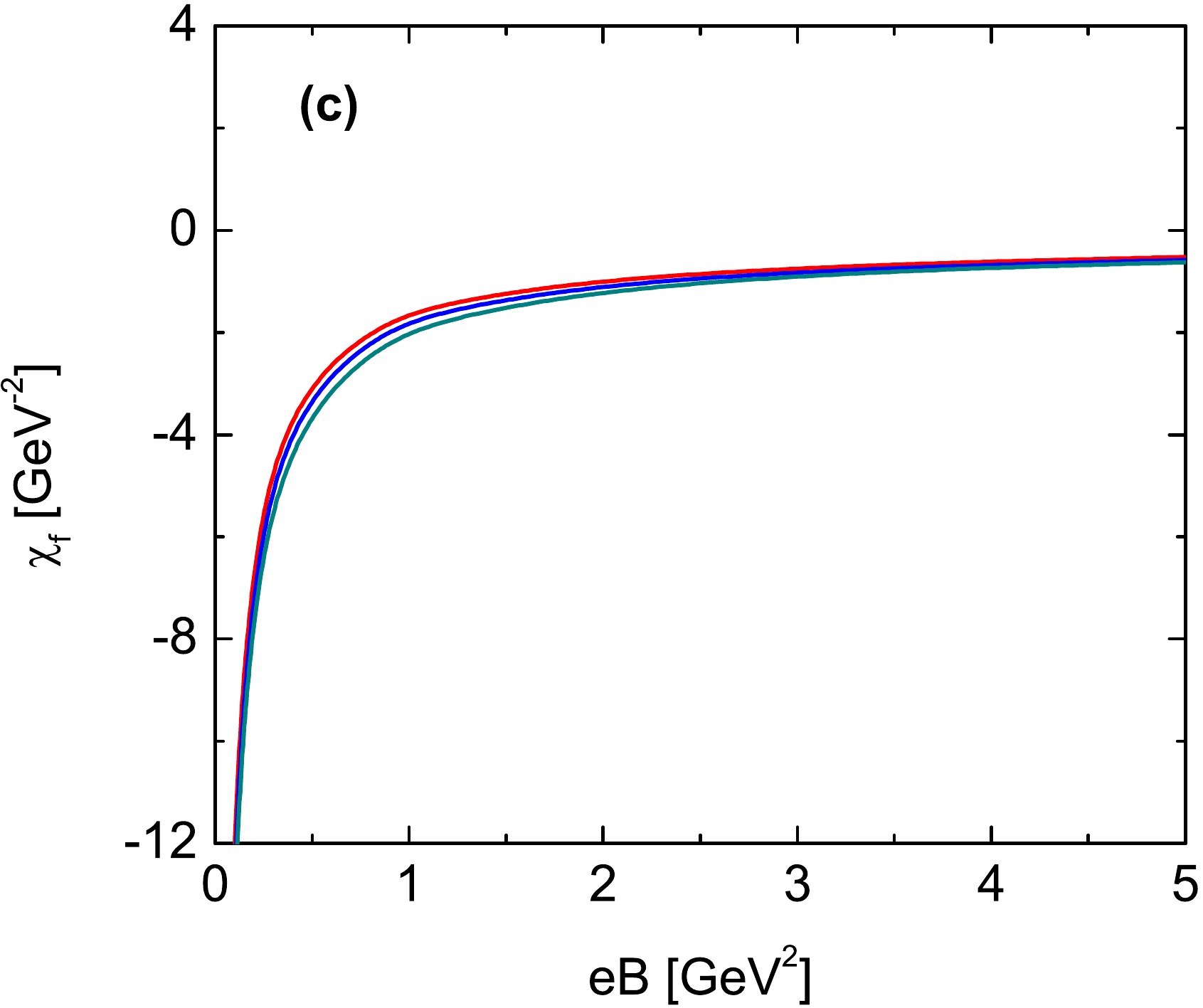}
\caption{(Color online)
The magnetic moment (left panel), magnetic polarization (middle panel) and magnetic susceptibility (right panel) of quarks,
as functions of the magnetic field $eB$, respectively.
}
\label{fig3}
\end{center}
\end{figure*}

\begin{figure*}[tp]
\begin{center}
\includegraphics[width=0.50\textwidth]{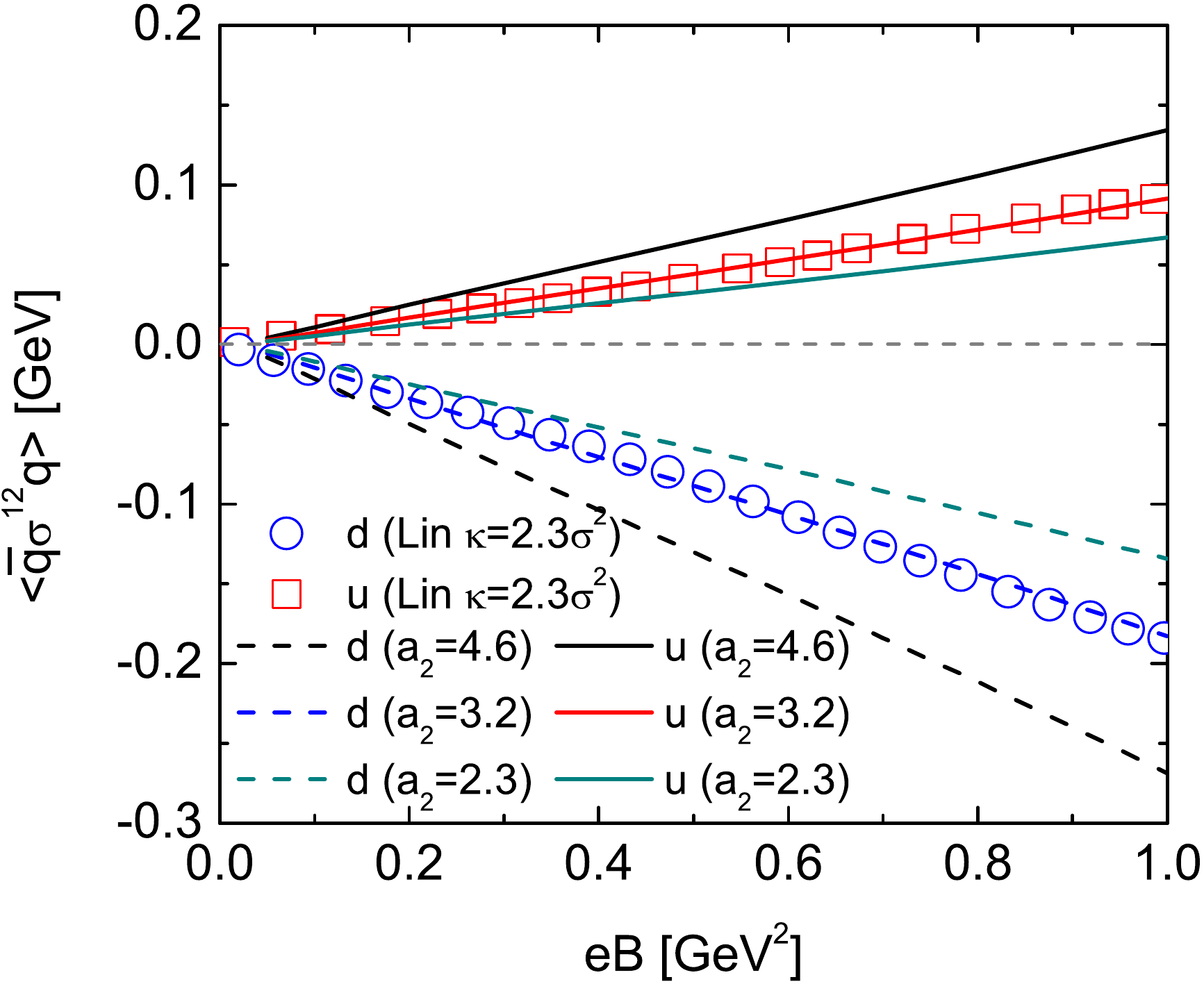}
\caption{(Color online)
The spin polarization of quarks, as a function of the magnetic field $eB$. Comparison between this work (solid and dotted lines) and the tensor-type spin polarization (TSP) results (hollow points)~\cite{Lin:2022moq}.
}
\label{fig4}
\end{center}
\end{figure*}

\begin{figure*}[h]
\begin{center}
\includegraphics[width=0.400\textwidth]{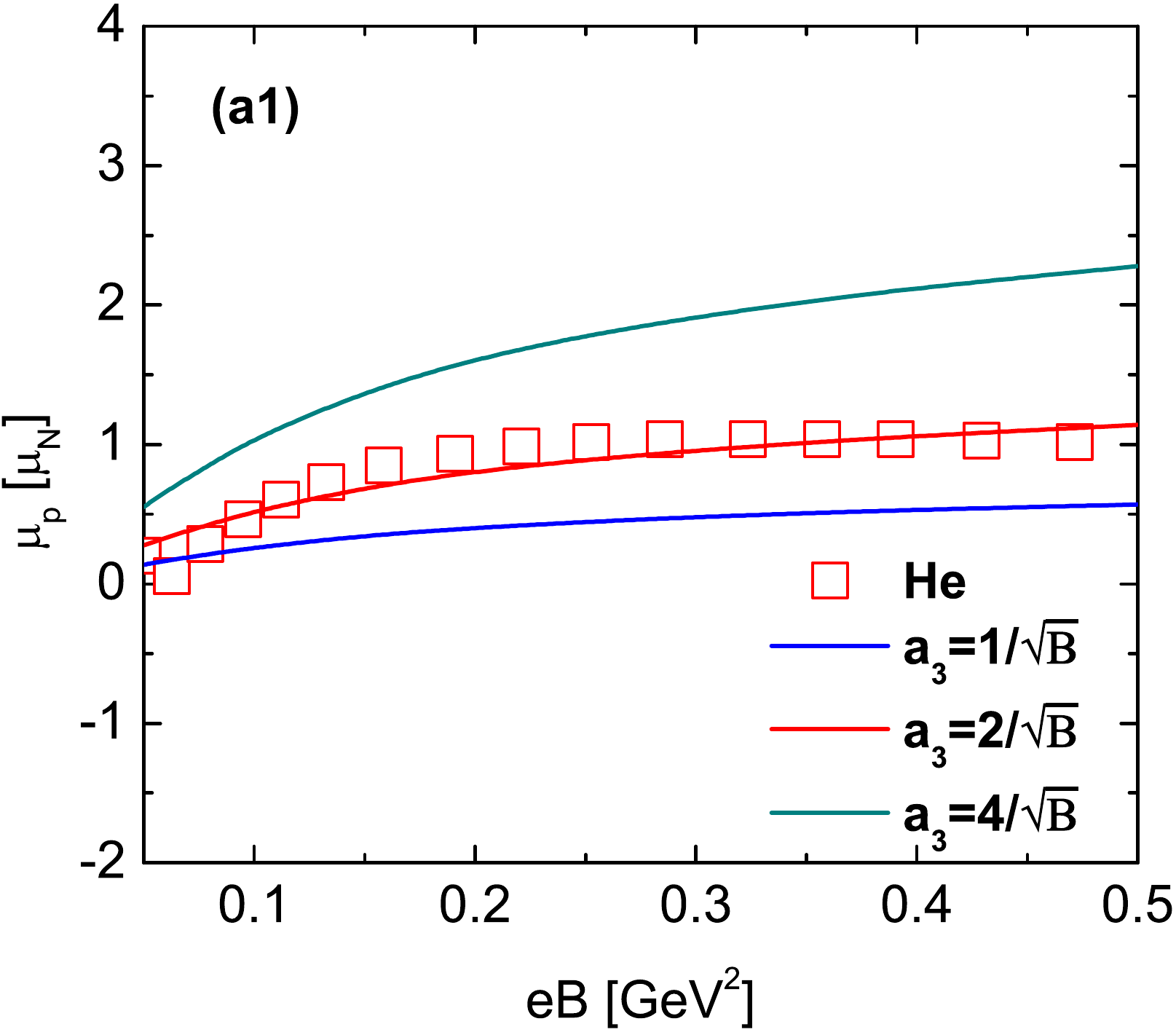}
\hspace{0.4cm}
\includegraphics[width=0.420\textwidth]{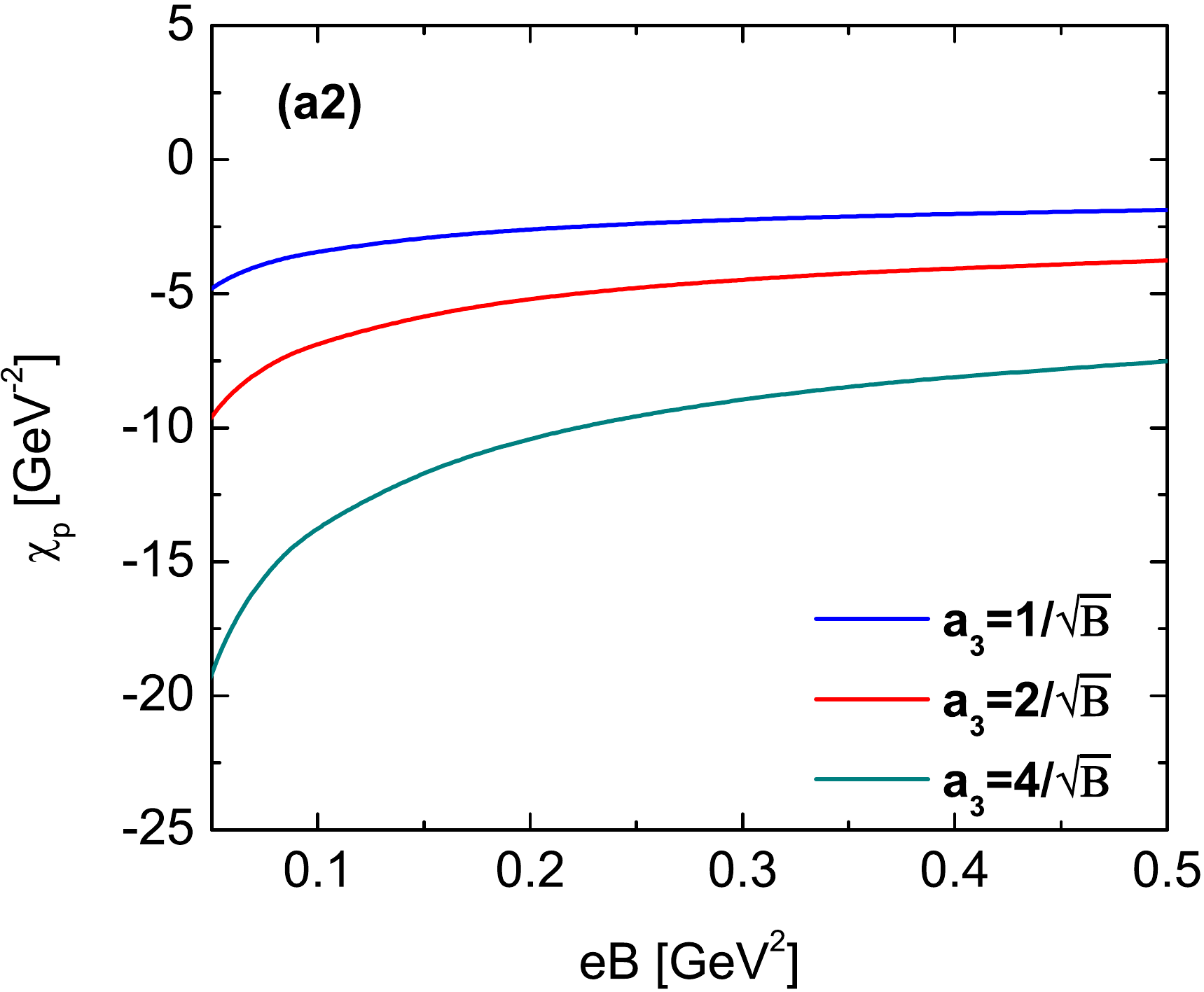}\\
\includegraphics[width=0.400\textwidth]{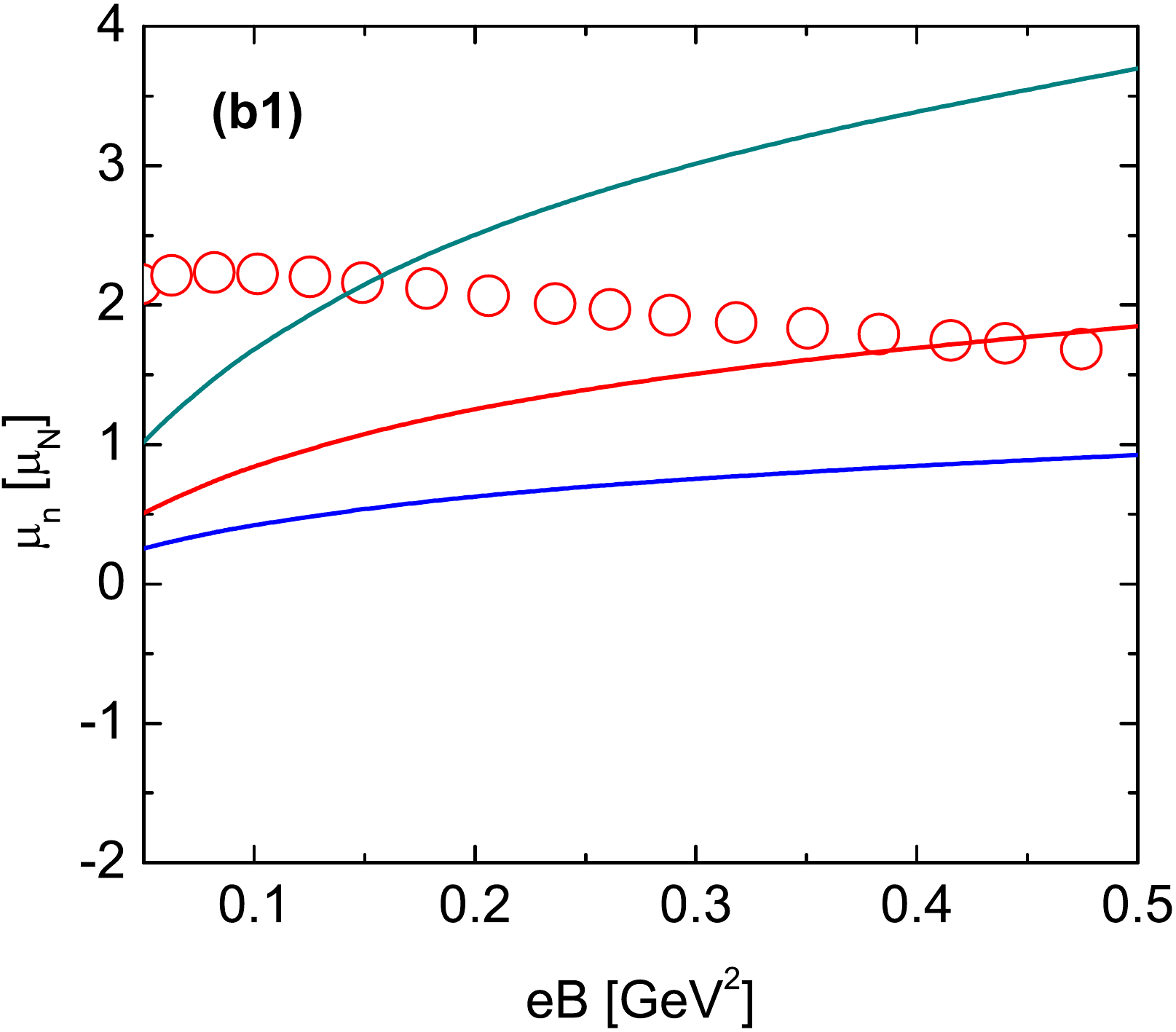}
\hspace{0.4cm}
\includegraphics[width=0.420\textwidth]{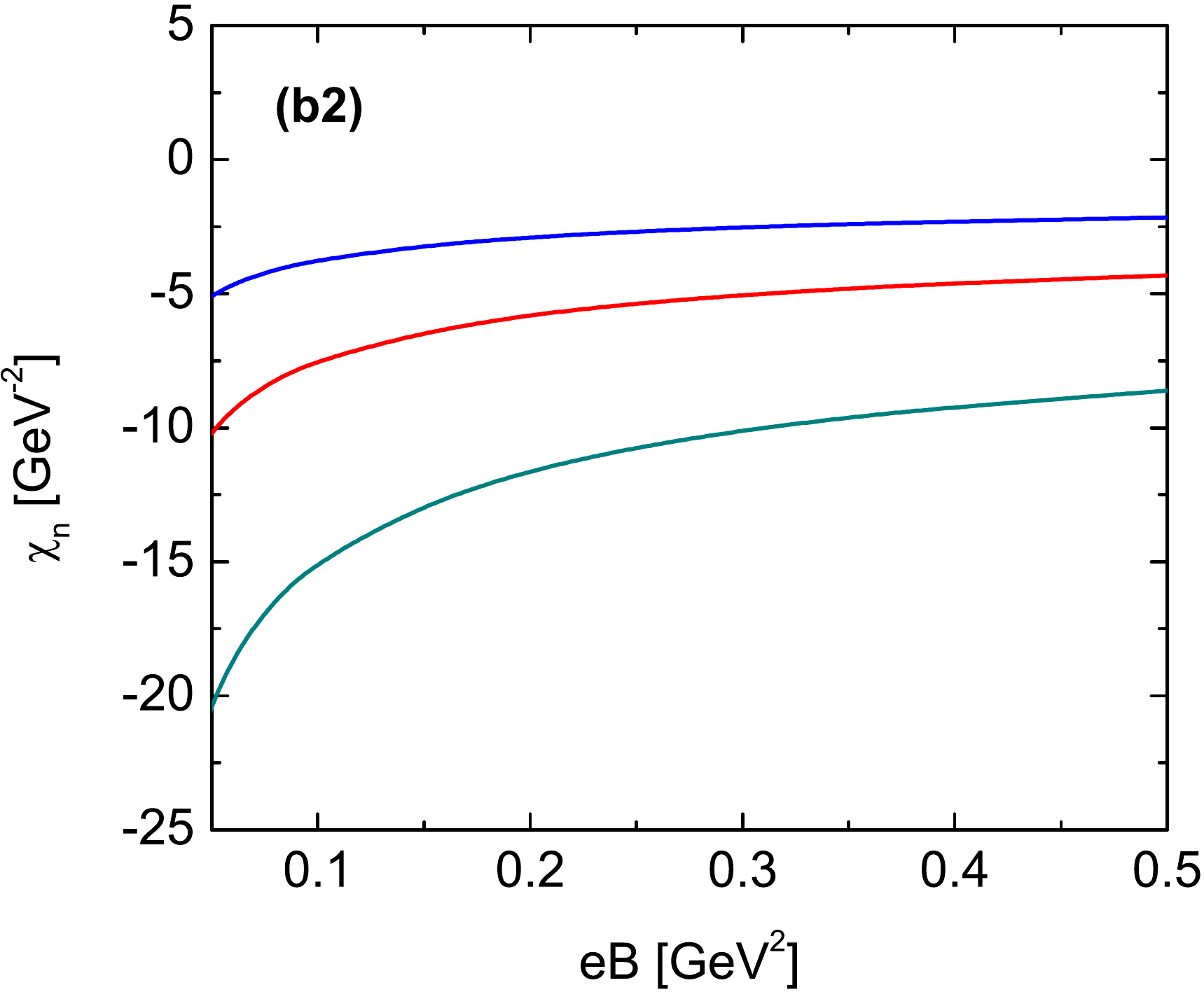}
\caption{(Color online)
The magnetic moments (left panels) and magnetic susceptibility (right panels) of nucleon, as functions of the magnetic field $eB$, respectively.
Up panels: results for proton. Down panels: results for neutron.
The magnetic moments of nucleon, are combined by the magnetic moments of quark [from \Fig{fig3} (a)].
The magnetic susceptibility of nucleon, are combined by the magnetic susceptibility of quark [from \Fig{fig3} (c)].
Comparison between our results (solid lines) and the Skyrme results (hollow points)~\cite{He:2017sms}.
}
\label{fig5}
\end{center}
\end{figure*}

This section presents the numerical results for the dynamically generated quark condensate,
magnetization phenomena, i.e., magnetic moment, magnetic polarization, spin polarization,
and magnetic susceptibility of a quark in the external magnetic field at zero temperature.
We also present the numerical results for the magnetic moments and magnetic susceptibility of a nucleon.

First, we display the quenched dressing functions $A_{\parallel}$
and $A_{\perp}$ (at the second lowest Landau level, $L=1$),
as functions of the momentum $p_{\parallel}^{2}$ for $m_{u}$=6 MeV at $eB$=1.0~GeV$^{2}$ in \Fig{fig1}, respectively.
From \Fig{fig1}, we can see quite clearly that, when the momentum is greater than 6 GeV$^{2}$,
the quenched dressing functions $A_{\parallel}$ and $A_{\perp}$ coincide with
each other, they are almost the same. However, for the momentum is less than about 6 GeV$^{2}$,
they become dramatically different.
In the mid-momentum region, one finds a characteristic bump in the dressing function.
The peak behaviour in the mid-momentum region (about $p_{\parallel}^{2}\approx $0.6 GeV$^{2}$)
can be understood as magnetic field breaking the Euclidean $O(4)$ symmetry to
an $O_{\parallel}(2)~\bigotimes~O_{\perp}(2)$~[as $p=(p_{\parallel},p_{\perp})=(p_{0},0,\sqrt{2|eB|L},p_{3})$ used in this paper].
Such a peak distribution is similar to the one found in Ref.~\cite{Watson:2014qge},
where a different approximation of the quark-DSE has been used.

Quark condensate is essential in describing nuclear matter and hadron physics at the lowest dimension.
The local quark condensate is given by
\begin{eqnarray}\label{qc:def101}
-\langle \bar{q}q\rangle &=& Z_{2}N_{c}\frac{eB}{2\pi^{2}}\sum_{L_{q}=0}^{\infty}\frac{\tau(L_{q})}{2}\int_{0}^{\infty}dq_{\parallel}q_{\parallel}
 \left[\frac{A_{0}(q)}{A_{0}^{2}(q)+A_{\parallel}^{2}(q)q_{\parallel}^{2}+ A_{\perp}^{2}(q)q_{\perp}^{2}}\right]\Bigg|_{L_{q}}.
\end{eqnarray}

Using the solutions of the quark's DSE in strong magnetic field at zero temperatures $A_{0}$, $A_{\parallel}$, and $A_{\perp}$,
we obtain the QCD vacuum properties in a strong magnetic field.
In the case $\langle \bar{q}q\rangle\neq0$, it means that the chiral symmetry is broken.

The QCD has printed out that the quark condensate involves the ultraviolet divergence at the zero-temperature part
and should be renormalized to be a finite quantity.
To eliminate the divergence arising in the quark condensate under external magnetic field,
one has introduced a dimensionless quantity of the subtracted quark condensate
in the lattice QCD~\cite{Bali:2012qqc,Ding:2022cca}
and NJL model [with the quark anomalous magnetic moment (AMM)~\cite{Kawaguchi:2022rot}].
Regarding the DSE work, we use a similar subtracted quark condensate and consider an additional parameter, $a_{1}/eB$, as
\begin{eqnarray}\label{qc:def102}
\frac{\left(\Sigma_{u}+\Sigma_{d}\right)}{2} &=& 1-\frac{a_{1}}{eB}\frac{m_{L}}{m_{\pi}^{2}f_{\pi}^{2}} \left\{\left[\langle \bar{u}u\rangle_{eB}-\langle \bar{u}u\rangle_{eB=0}\right]+\left[\langle \bar{d}d\rangle_{eB}-\langle \bar{d}d\rangle_{eB=0}\right]\right\},
\end{eqnarray}
where $m_{L}=6$~MeV, $m_{\pi}=138$~MeV and $f_{\pi}=92.4$~MeV.
The $a_{1}/eB$ parameter plays the part of the anomalous magnetic moment factor of the external magnetic-dependent chiral condensate
(one describes the chiral symmetry-breaking term),
which is compared with the AMM effect (noted as $\kappa_{u}$, $\kappa_{d}$) in NJL~\cite{Kawaguchi:2022rot}.
The quark condensates $\langle \bar{u}u\rangle$ and $\langle \bar{d}d\rangle$ are obtained from~\eq{qc:def101}.

In \Fig{fig2}, we show the magnetic field dependence of the subtracted quark condensate
for four different $a_{1}$ parameters, in comparison with the lattice QCD~\cite{Ding:2022cca} and the NJL (with/without AMM)~\cite{Kawaguchi:2022rot}.
The subtracted quark condensate monotonically grows with the increases of the magnetic field, shows the magnetic catalysis effect.
$(\Sigma_{u}+\Sigma_{d})/2$ depends on parameter $a_{1}$ and increases with the increase of $a_{1}$.
Our result for $a_{1}=0.04$ is in good agreement with the lattice QCD result and the NJL result(with the magnetic-dependent AMM).
In this work, we consider the quenched ladder approximation with constant coupling $g$,
so the constant coupling $g$ in some degree acts as a catalyzer for the chiral symmetry breaking under a magnetic field in the zero temperature.

Next, we investigate the property of magnetization phenomena in detail.
The novel domain of magnetic moment interactions in external fields,
which has been recently formulated, is promising in enhancing the understanding of the physics of plasmas.

The knowledge about the strength of the induced magnetization
is provided by the magnetic moment and susceptibility.
The magnetic moment of a quark is given by~\cite{Bali:2012mso,Mueller:2014dqm}
\begin{eqnarray}\label{sp:def101}
\langle \sigma^{12}\rangle_{f} &=& \mu_{f}=Z_{2}N_{c}\frac{eB}{2\pi^{2}}\int_{0}^{\infty}dq_{\parallel}q_{\parallel}
 \left[\triangle sgn(eB)\frac{A_{0}(q)}{A_{0}^{2}(q)+A_{\parallel}^{2}(q)q_{\parallel}^{2}}\right].
\end{eqnarray}

Using values for the magnetic moment, the zero-temperature magnetic polarization and susceptibilities are calculated as
\begin{eqnarray}\label{sp:def102}
P_{f} &=& \frac{\langle \sigma^{12}\rangle_{f}}{\langle \bar{q}q\rangle},~~~~\chi_{f} \approx \frac{a_{2}P_{f}}{eB},
\end{eqnarray}
where $a_{2}$ plays the part of a magnetic rotation effect,
i.e., $a_{2}\approx\langle \bar{q}\sigma^{12}q\rangle/\langle \sigma^{12}\rangle_{f}$,
which is compared to the definition in Ref.~\cite{Watson:2014qge}.

We now investigate the results of magnetized QCD, i.e.,
magnetic moment $\langle \sigma^{12}\rangle_{f}$, magnetic polarization $P_{f}$ and magnetic susceptibility $\chi_{f}$.
These magnetization phenomena of quark, as functions of the magnetic field $eB$, are shown in \Fig{fig3}, respectively.
\Fig{fig3} highlights that both the magnetic moment and magnetic polarization increase as the magnetic field increases,
while the magnetic susceptibility decreases its absolute magnitude with an increasing magnetic field.
The magnetic moment, magnetic polarization, and magnetic susceptibility results of the quarks also depend on the quark's mass.
Here, we discuss the three quantities presented in the figures:
(i) The quark model~\cite{Casu:1997bmm} has fitted out a negative magnetic moment of $d$ quark, $\langle \sigma^{12}\rangle_{d}$,
while we obtain a positive value (left panel). Similarly positive values have also been pointed in the NJL model~\cite{Kawaguchi:2022rot,Felipe:2008msq,Mei:2020ice,Farias:2022eot,Lin:2022moq}.
(ii) The lattice QCD has pointed out that saturation and magnetic polarization $P_{f}\rightarrow$~1
as $eB\rightarrow$ a huge value ~\cite{Buividovich:2010cmo},
while $P_{f}>$~1 as $eB\rightarrow$ a huge value shown in this work (middle panel).
In fact, dressed quarks possess a large, dynamically generated anomalous chromomagnetic moment,
which produces an equally large anomalous electromagnetic moment that
impact on the nucleon magnetic form factors~\cite{Chang:2011tds}.
(iii) The lattice QCD noted that sign conventions $\chi_{f}>0$ correspond to paramagnetism
(in which the exposure to the background field is energetically favorable)
and $\chi_{f}<0$ to diamagnetism (which repels the external field) by using the equivalence.
The results of $\chi_{f}<0$ to diamagnetism are also shown in the tensor-type spin polarization (TSP) with AMM~\cite{Lin:2022moq},
while $\chi_{f}>0$ without TSP. Our results (right panel) show a negative $\chi_{f}$, as a consequence of the negative $P_{f}$.

For a general background field $eB$, the spin polarization of a quark
can be extracted from the quark condensate term through~\cite{Bali:2012mso,Lin:2022moq,Bali:2020mso}
\begin{eqnarray}\label{sp:def103}
\langle \bar{q}\sigma^{12}q\rangle &=& Q_{f}\cdot eB\cdot \langle \bar{q}q\rangle\cdot\chi_{f} \approx Q_{f}\cdot eB\cdot \langle \bar{q}q\rangle\cdot \frac{a_{2}P_{f}}{eB} =a_{2}Q_{f}P_{f}\cdot \langle \bar{q}q\rangle,
\end{eqnarray}
where $Q_{f}$ is the charge of quarks, and $\langle \bar{q}q\rangle$ is obtained from~\eq{qc:def101}.
One might suspect that $\chi_{f}$ and $\langle \bar{q}\sigma^{12}q\rangle/eB$ are not completely unrelated.
Indeed, as Bali and his collaborators have discussed in Ref.~\cite{Bali:2012mso,Bali:2020mso},
$\langle \bar{q}\sigma^{12}q\rangle/eB$ represents the contribution of the spin of quark flavor to
the total magnetic susceptibility (such total magnetic susceptibility is composed of spin-related
and orbital angular momentum-related contributions).
Here, we consider the spin-related quantity using the parameter $a_{2}$.
Note that, the sub-index $f$ in \eq{sp:def101}-\eq{sp:def103} is the flavor of quark.

The spin polarizations of quark, as functions of the magnetic fields $eB$ and for various $a_{2}$, are shown in \Fig{fig4}.
These spin polarizations of $u/d$ quark can be obtained from \eq{sp:def103}.
The figure infers that the spin polarizations are approximately linear for magnetic field $eB$,
and its absolute magnitude increase as the $eB$ increases (while vanishes for $eB$= 0, as it should).
As $eB$ increases, due to the difference between the electric charges, the spin polarizations of $u/d$ quark develop an $a_{2}$-dependence.
Comparisons between our results (solid and dot lines) and the TSP with AMM results (hollow points)~\cite{Lin:2022moq},
are shown in the figure.
Our results (with $a_{2}=3.2$) agree with the TSP with AMM results.
Note that $\langle \bar{q}\sigma^{12}q\rangle$ of our results and
the TSP with AMM results (hollow points)~\cite{Lin:2022moq}
are positive, while is negative in Ref.~\cite{Bali:2012mso}.
It can be understood that in this work, the spin polarization definition is along the direction of the external magnetic field.

The study on the anomalous magnetic moment of baryons, i.e., proton and neutron, has been conducted for many years~\cite{Zyla:2020rop}.
Such an anomalous magnetic moment of nucleon has been considered using the modified 'weak' field expansion of
the fermion propagator having non-trivial correction terms for charged and neutral particles.
The DSE approach~\cite{Cloet:2009son,Bednar:2018nqd} was used to calculate and predict the nucleon's electromagnetic form factors
and parton distribution functions.
Furthermore, the constituent quark model has pointed out that the magnetic moment of a proton
is the vector sum of the magnetic moments of its three constituent quarks
$\vec{\mu}=\vec{\mu}_{1}+\vec{\mu}_{2}+\vec{\mu}_{3}$~\cite{Donoghue:2014dos}
(where the sub-index $1,2,3$ correspond to the $u,d,s$ quarks).
In this work, we introduce the properties of the magnetic moment of nucleons in strong magnetic field at zero temperature.
To estimate these properties, we follow the idea of the constituent quark model~\cite{Donoghue:2014dos}. Here is
\begin{eqnarray}\label{sp:def104}
\mu_{p} &=& a_{3}\left[\frac{4}{3}\mu_{u}-\frac{1}{3}\mu_{d}\right] =a_{3}\left[\frac{4}{3}\langle \sigma^{12}\rangle_{u}-\frac{1}{3}\langle \sigma^{12}\rangle_{d}\right],~\nonumber\\
\mu_{n} &=& a_{3}\left[\frac{4}{3}\mu_{d}-\frac{1}{3}\mu_{u}\right] =a_{3}\left[\frac{4}{3}\langle \sigma^{12}\rangle_{d}-\frac{1}{3}\langle \sigma^{12}\rangle_{u}\right].
\end{eqnarray}
where $\mu_{u}=\langle \sigma^{12}\rangle_{u}$,~and $\mu_{d}=\langle \sigma^{12}\rangle_{d}$
are the magnetic moments of quark [obtained from~\eq{sp:def101}].
The sub-index $p$,~$n$ are noted as proton and neutron, respectively.
Besides, the parameters $a_{3}$ plays the part of spin and total orbital angular momentum effects~\cite{Casu:1997bmm} of the twisted shape of nucleons,
where are compared to Ref.~~\cite{He:2017sms}.

Similarly, we calculate the magnetic susceptibility of nucleon as in the constituent quark model,
\begin{eqnarray}\label{sp:def105}
\chi_{p} &=& a_{3}\left[\frac{4}{3}\chi_{u}-\frac{1}{3}\chi_{d}\right] \approx a_{3}\left[\frac{4}{3}\frac{a_{2}P_{u}}{eB}-\frac{1}{3}\frac{a_{2}P_{d}}{eB}\right],~\nonumber\\
\chi_{n} &=& a_{3}\left[\frac{4}{3}\chi_{d}-\frac{1}{3}\chi_{u}\right] \approx a_{3}\left[\frac{4}{3}\frac{a_{2}P_{d}}{eB}-\frac{1}{3}\frac{a_{2}P_{u}}{eB}\right].
\end{eqnarray}

The magnetic moments (left panels) and magnetic susceptibility (right panels) of a nucleon,
as functions of the magnetic field $eB$, for three different $a_{3}$ parameters, are shown in \Fig{fig5}, respectively.
The results of the proton are shown in the upper panels, and the results of the neutron are shown in the lower panels.
Such magnetic moments of nucleon are combined by
the magnetic moments of quark from ~\eq{sp:def101} [as in \Fig{fig3} (a)].
Additionally, the magnetic susceptibility of nucleons is combined with
the magnetic susceptibility of quark from~\eq{sp:def102} [as in \Fig{fig3} (c)].
We compare our results (solid lines) of $\mu_{p}$ and $\mu_{n}$
to the Skyrme model results (hollow points)~\cite{He:2017sms},
which are presented in \Fig{fig5}.
Our results agree with the absolute magnitude of the Skyrme result,
presenting differences for $eB=0$ (or $eB$ sufficiently low),
and also for the $eB$-dependent magnetic moment of the neutron.
Note that we consider $\mu_{p}$~(and $\mu_{n}$)~$>0$, while $\mu_{p}$~(and $\mu_{n}$)~$<0$
in the Skyrme model at $eB\geq$ 0.062~GeV$^{2}$.
Such different results reveal a linearly magnetic dependence $\mu$ of a quark in the DSE definition~\cite{Mueller:2014dqm},
while the mass and magnetic-dependent $\mu$ of the twisted shape of nucleon are considered
in an effective polarization space by the Skyrme model~\cite{He:2017sms}.
Note that a prediction for the ratio of $\mu_{p}/\mu_{n}\rightarrow 1$ when $eB\rightarrow \infty$
(not shown in the figure) agrees with the prediction of the Skyrme model~\cite{He:2017sms}.
Besides, the prediction for $\chi_{p/n}<0$ corresponds to the spin-diamagnetic behavior of the nucleon,
which is combined with the quarks.
\section{Summary}
\label{sec:sum}
In summary, we study on the condensate and magnetization phenomena, i.e., the magnetic moment,
magnetic polarization, spin polarization, and magnetic susceptibility in strong external
magnetic field based on the framework of DSE calculations. The results shown that
such condensate and magnetization phenomena are dependent on the magnitude of magnetic field.

(I) With the solutions of the quarks DSE,
the magnetic field dependence of the subtracted quark condensate for four different $a_{1}$ parameters are obtained.
We find that the condensate monotonically grows with the increases of the magnetic field, shows the magnetic catalysis effect.
Our result for $a_{1}=0.04$ is in good agreement with the lattice QCD result and the NJL result (with the magnetic-dependent AMM).
In this work, we consider the quenched ladder approximation with constant coupling $g$,
so the constant coupling $g$ in some degree acts as a catalyzer for the chiral symmetry breaking
under a magnetic field in the zero temperature.
Actually Refs.~\cite{Farias:2014prc,Ferrer:2015qaa,Farias:2017tei,Tavares:2021nsm} found that the inverse magnetic catalysis phenomenon
is associated to the effect of the magnetic field on the QCD coupling constant.
Hence, it would be valuable to reevaluate the effect of the magnetic field on the QCD coupling constant in the future.

(II) It is also shown that the solutions for the magnetization phenomena, i.e., the magnetic moment,
magnetic polarization, spin polarization, and magnetic susceptibility, as functions of magnetic field.
Besides, it is found that in a strong magnetic field the state with the positive magnetic moment of $u/d$ quark,
one can be realized as results from Landau levels, while is difference of the results of the quark model
(it consider that $u$ quark in a proton contributes with a positive anomalous magnetic moment and $d$ quarks with a negative value).
The magnetic susceptibilities $\chi_{f}<0$ reveal a spin-diamagnetic behavior at zero temperature,
which is similar to the result of lattice QCD.

(III) We also conclude that magnetic dependent $\mu_{p/n}$ and $\chi_{p/n}$
can be combined with the quark results using the constituent quark model.
These predictions for $\mu_{p/n}>0,~\chi_{p/n}<0$ correspond to the spin-diamagnetic behaviors for the nucleon.

\section*{Acknowledgements}
D.-X. Wei has been supported by the National Natural Science Foundation of China Grant No.~12105057,
the Youth Program of Natural Science Foundation of Guangxi (China) Grant No.~2019GXNSFBA245080.
L.-J. Zhou has been supported by the National Natural Science Foundation of China Grant No.~11865005.

\end{document}